\documentclass[prl,reprint,amsmath,amssymb,twoside,twocolumn]{revtex4-2}
\usepackage{graphicx}
\usepackage{bbm}
\usepackage{xcolor}
\usepackage[T1]{fontenc}
\usepackage{tikz}
\usetikzlibrary{arrows,arrows.meta,calc}
\usetikzlibrary{decorations.pathreplacing}

\definecolor{cobalt}{rgb}{0.0, 0.28, 0.67}
\usepackage[colorlinks = true,
linkcolor = cobalt,
urlcolor  = cobalt,
citecolor = cobalt,
anchorcolor = cobalt]{hyperref}

\newcommand{\eps}{\varepsilon}
\newcommand{\nx}{n(x)}
\newcommand{\n}{n}
\newcommand{\g}{\gamma}%division rate
\newcommand{\de}{d}%%death rate
\newcommand{\cvd}{CV^2_{d}}
\newcommand{\cvg}{CV^2_{\g}}
\newcommand{\md}{\mu_{\de}}
\newcommand{\mg}{\mu_{\g}}
\newcommand{\s}{s}
\newcommand{\genfun}{\mathcal{Z}}
\newcommand{\prfun}{\mathcal{P}}
\newcommand{\Pid}{\Pi_d}
\newcommand{\Pig}{\Pi_{\g}}
\newcommand{\Pii}{\Pi}
\newcommand{\psurv}{p_\text{surv}}

\newcommand{\delt}{\delta_{\bullet}}
\newcommand{\Zon}{\genfun_\text{on}}
\newcommand{\Zoff}{\genfun_\text{off}}

\newcommand{\changes}[1]{{#1}}
\newcommand{\cchanges}[1]{{#1}}

\newcommand{\App}{SM}

\begin{document}
\title{\cchanges{Survival resonances during fractional killing of cell populations}}
\author{Francesco Puccioni}
\thanks{These two authors contributed equally}
\author{Johannes Pausch}
\thanks{These two authors contributed equally}
\author{Paul Piho}
\author{Philipp Thomas}
\affiliation{Imperial College London, South Kensington, London SW7 2AZ, United Kingdom}

\begin{abstract}
Fractional killing in response to drugs is a hallmark of non-genetic cellular heterogeneity. Yet how individual lineages evade drug treatment, as observed in bacteria and cancer cells, is not quantitatively understood. We study a stochastic population model with age-dependent division and death rates, allowing for persistence. In periodic drug environments, we discover peaks in the survival probabilities at division or death times that are multiples of the environment duration. The survival resonances are unseen in unstructured populations and are amplified by persistence.
\end{abstract}

\maketitle

\begin{figure}
\includegraphics[width=\columnwidth]{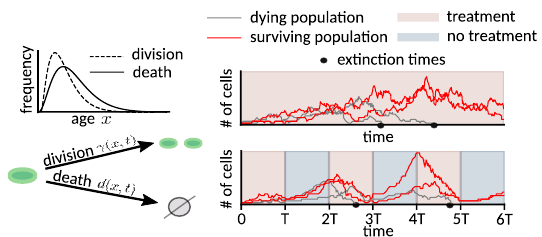}
\caption{Illustration of age-dependent population dynamics. Top left panel: the distribution of division and death ages. Bottom left: cells divide and die depending on age and environment. Top right: constant environment. Bottom right: periodic on- and off-switching of a treatment. Some cells survive treatment and persist, fractional killing occurs.}\label{fig1}
\end{figure}

Under adverse conditions such as repeated drug treatments, most cells in a population die while few cells survive -- a phenomenon called \textit{fractional killing} \cite{Balaban2004,shaffer2017rare,Kessler2022}. The short timescale of drug exposure often excludes the evolution of drug resistance but requires non-genetic mechanisms underlying fractional killing, which are still not fully understood \cite{Rotem2010,Feng2014,Oren2021}. A well-accepted view is that heterogeneous survival arises from fluctuations in intracellular pathways influencing cell division, growth and apoptosis in coordination with the cell cycle \cite{Spencer2009,Paek2016,granada2020effects}. 
Recent advances in single-cell imaging allow tracking heterogeneity in individual lineages that can drive insights into persistence against antimicrobial or anti-cancer treatments aided by quantitative stochastic models.
Cellular heterogeneity in division times, as observed in experiments, can be described through age-structured branching processes \cite{Harris1964,Kimmel2015,jafarpour2018bridging,jafarpour2019cell,nozoe2020cell,levien2020large,piho2024}.
Most of these models operate on a mean-field level as they often implicitly assume cell survival or large populations, and therefore cannot explain the fractional killing. Here, we provide a quantitative model of fractional killing for populations established from a single ancestor by analysing an age-structured branching process of cell division and death in time-dependent environments \cite{Ballweg2017,Chakrabarti2018,genthonPRX2023} (Fig.~\ref{fig1}). As fractional killing is the survival of cells despite long or repeated exposure to adverse conditions, we focus on long-term, i.e.~infinite-time, survival. We will refer to it simply as survival unless stated otherwise. 

In this letter, we present several results towards understanding fractional killing as a stochastic phenomenon that emerges due to heterogeneity and time-dependent treatments. To achieve this, \changes{we consider the survival probability in constant environments and quantify fractional killing with persistence, where cells fail to die or divide.} Next, we generalise the age-dependent branching model of Bellman and Harris \cite{Harris1964}, allowing arbitrary age-dependence of division and death rates in periodic environments. Here, we discover that periodic treatment settings can lead to resonance phenomena in the survival probability \changes{that are amplified by persistence}. Our results reveal a complex dependence of resonances on division or death times and the environment that could be observed for cell-cycle dependent drugs such as chemotherapeutic drugs or antibiotics \cite{sun2021influence,kohanski2010antibiotics,Lock2008,Cutty2024}.

\textit{Model:} We consider a population of cells that age deterministically and are subject to spontaneous division or death events in changing environments. At time $t$, a cell $\s(x)$ with age $x$ divides into two newborns with rate $\g(x,t)$ or dies with rate $\de(x,t)$: 
\begin{equation}
\label{eq::stoichiometric_birth_death}
\s(x)\xrightarrow{\g(x,t)} 2\s(0),\ \
\s(x)\xrightarrow{\de(x,t)} \varnothing 
\end{equation}  
Here, the age $x$ labels the time since the last division.
The age-dependence of the rates means that the division and death time distributions of reactions~\eqref{eq::stoichiometric_birth_death} are non-exponential and cell cycle-dependent as observed in experiments \cite{Chakrabarti2018,gross2023analysis}. Together with the time-dependence of these rates, they determine not only cell fate but ultimately the survival probability $\psurv$ of the entire population. We follow the evolution of the cell population with a function $\n$ where $\nx$ is the density of cells with age $x$. Starting at time $t=0$ with a single ancestor of age $x_0$, the probability of observing a specific density is denoted by the functional $\prfun[\n,t|\delta_{x_0}]$ such that $\psurv(x_0)=1-\lim\limits_{t\rightarrow\infty} \prfun[0,t|\delta_{x_0}]$. $\prfun$ evolves according to a functional Master equation \cite{vanKampen1977,O'Dwyer2009}: 
\begin{align}\label{eq: master_equation}
	&\frac{\partial \prfun[\n,t|\delta_{x_0}]}{\partial t}=\int_0^{\infty} \!\! {\rm{d}}x\biggl(\g(x,t)(\eps^{+1}_x\eps^{-2}_0-1)\nx+\notag\\
	&\quad\de(x,t)(\eps^{+1}_x-1)\nx+\frac{\delta }{\delta \nx}\partial_x\nx\biggr) \prfun[\n,t|\delta_{x_0}],
\end{align}
\cchanges{where the first two terms describe age-dependent division and death processes and the third term is due to the increase of age $x$ with time $t$ leading to an effective drift term.} 
Here, the step operator $\eps^{\pm m}_x$ shifts the argument of any functional $\mathcal{F}$: $\eps^{\pm m}_x \mathcal{F}[n]= \mathcal{F}[\n\pm m\delta_x]$ by a Dirac-$\delta$ function $\delta_x$. For example, \cchanges{$\eps^{-2}_0\eps^{+1}_x$} stands for dividing a cell with age $x$ into two cells of zero age. \cchanges{The mean-field of the master equation is the McKendrick-von Foerster model (SM \ref{SM::growthrate},  \cite{VonFoerster1959}).}

\begin{figure}
\includegraphics[width=0.47\textwidth]{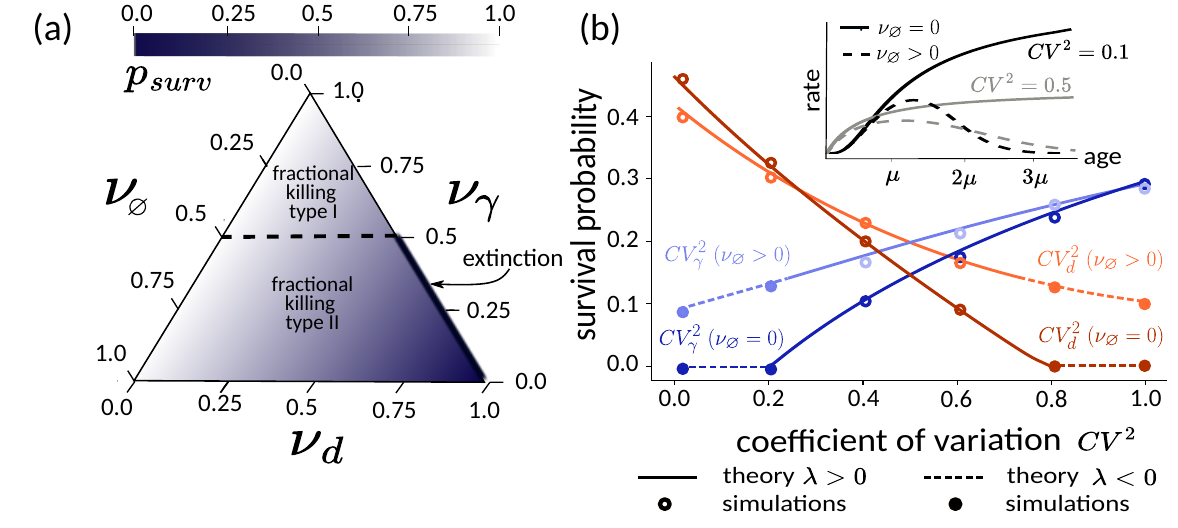}
 \vspace{-0cm}
 \caption{Noise-induced transitions in constant environments. 
 (a) Phase diagram showing the regions of extinction (black line), fractional killing type I ($\lambda>0$) and type II ($\lambda<0$). 
 (b) Survival probability shows transitions from extinction to fractional killing ($\nu_\varnothing=0$), and type I to type II killing ($\nu_\varnothing>0$) as the coefficient of variation in division or death times is varied, $\cvg$ or $\cvd$. $\Gamma$-distributed times assumed with $\md=1$, $\mg=0.9$, $w_{\g}=w_d=\{0.8,1.0\}$, $\cvd=\cvg=0.5$ unless stated otherwise.}
  \label{fig2}
\end{figure}

\textit{Constant environment: }
\changes{
In our model, cells can either divide, die or persist, which occurs with probabilities $\nu_{\gamma}$ (division), $\nu_d$ (death) and $\nu_{\varnothing}=1-\nu_{\gamma}- \nu_d$ (persistence). Persistence is an intrinsic property of our age-structured model and does not occur for constant rates. To see this, we note that any rate function can be written as $y(x)=\frac{w_y f_y(x)}{w_y\int_x^{\infty}f_y(u)\mathrm{d}u+(1-w_y)}$ \label{page:eq_rates_with_peristence} using the event time distribution $f_{y}(x)$ and the probability of that event $w_y$ \footnote{The rate is constant only for the exponential distribution with $w_y=1$.}. 
This implies that the persistence probability is $\nu_\varnothing=(1-w_{d})(1-w_{\g})$ and $\nu_{\gamma}=\int_0^{\infty}{\textrm d}u\gamma(u)e^{-\int_0^u{\textrm d}y(\gamma(y)+d(y))}$ and $\nu_d=\int_0^{\infty}{\textrm d}u\de(u)e^{-\int_0^u{\textrm d}y(\gamma(y)+d(y))}$ (\App~\ref{app:characteristics}).

The extinction probability is the fixed point of the generating function of offspring (over one generation) $z(h)=\nu_d+\nu_{\gamma} h^2$, which yields:
\begin{align}\label{pext}
    \psurv(0)=
    1-\frac{1-\sqrt{1-4 \nu_d \nu_\gamma}}{2 \nu_\gamma},
\end{align}
for a lineage starting with zero age. 
When $\nu_\varnothing=0$, Eq.~\eqref{pext} demonstrates a second-order phase transition in the survival probability between a subcritical phase ($\frac{\nu_d}{\nu_{\gamma}}>1$) with almost sure extinction and a supercritical phase ($\frac{\nu_d}{\nu_{\gamma}}<1$) with positive survival probability, similar to the age-independent processes \cite{Kendall1948b,Garcia-Millan2018}. When $\nu_\varnothing>0$, cells survive also below the transition point. The second-order transition is replaced by a transition \cchanges{from an exponentially growing population to a finite pool of persister cells on average, which is indicated by a change of sign in growth rate. The growth rate $\lambda$ quantifies the exponential increase/decrease of the proliferating subpopulation and the timescale of persister formation (SM \ref{SM::growthrate}). It} is the solution to the Euler-Lotka equation $1=2\int_0^\infty\textrm{d}u e^{-\lambda u}\gamma(u)e^{-\int_0^u{\textrm d}y(\gamma(y)+d(y))}$. While the transition in survival probability occurs only when $\nu_{\gamma}=\nu_d=\frac{1}{2}$, the transition of growth rate occurs when $\nu_{\gamma}=\frac{1}{2}$ (Fig.~\ref{fig2}a) regardless of $\nu_\varnothing$. These lines separate the phases between extinction ($p_\text{surv}=0$, $\lambda<0$) and two fractional killing phases that we call type I ($p_\text{surv}>0$, $\lambda>0$) and type II  ($p_\text{surv}>0$, $\lambda<0$).}
We computed the survival probability for $\Gamma$-distributed division and death times, which allows us to vary the coefficients of variation of division and death times at constant mean times ($\mg$ and $\md$). As shown in Fig.~\ref{fig2}b, the survival probability increases with noise in division times ($\cvg$) and decreases with noise in death times ($\cvd$). Interestingly, \changes{for $\nu_\varnothing=0$}, this dependence induces a transition from complete to fractional killing as division noise increases (Fig.~\ref{fig2}b), and the reverse transition is observed as death timing noise increases. \changes{The transition disappears for $\nu_\varnothing>0$ and is replaced with a transition from type I to type II fractional killing as the survival probability crosses $\sqrt{2\nu_\varnothing}$. At this point, the dynamics change from growing to decaying lineages due to the effect of noise. A similar dependence is observed for log-normal distributed times (SM Fig.~\ref{sfig:lognormal}). We expect intricate parametric dependence to emerge for more complex distributions, which is comprehensively understood using the phase diagram (Fig.~\ref{fig2}a). }

\begin{figure*}[t!]
\includegraphics[width=\textwidth]
{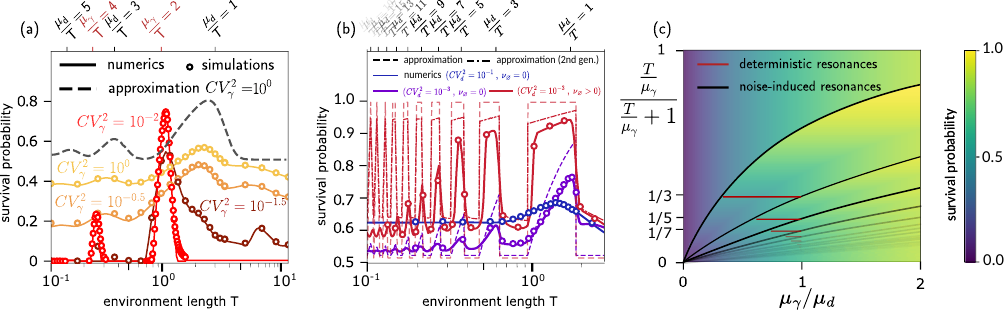}
 \vspace{-0.5cm}
  \caption{Survival resonances under periodic on-off treatments of length $T=T_\text{on}=T_\text{off}$ \changes{($x$-axis (a) and (b) and $y$-axis in (c) on log-scale)}.  (a): Survival probability ($\psurv(0)$) shows a crossover from deterministic (red) to noise-induced resonances (yellow and orange) as division noise ($\cvg$) increases.  (b): Noise-induced survival resonances increase with decreasing noise in death time (blue and violet) \changes{and are amplified by persistence ($\nu_\varnothing>0$, red) in agreement with the approximations (Eq.~\eqref{pext} with \eqref{eff_death_1st} and 2nd generation, SM Eq.~\eqref{eq:2nd_gen_approx}, $CV_d^2=0$).} Theoretical locations of deterministic (red, Eq.~\eqref{det_res}) and noise-induced resonances (black ticks, Eq.~\eqref{eq_noise-induced_resonance}) are shown as top ticks. (c): Phase diagram of survival probability peaks in rescaled $T$-$\md/\mg$ space with heatmap of the approximation (Eq.~\eqref{pext} with \eqref{eff_death_1st}) \changes{for $\nu_\emptyset=0$}. Division and death times are $\Gamma$ distributed with parameters $\mg=2$, $\md=1.8$, $\cvd=10^{-2}$, $w_\gamma=w_d=1$ in (a) and $\mg=1.8$, $\md=2$, $\cvg=1$, $w_\gamma=w_d=1$ in (b,c) or $0.99$ in (red, b). }\label{fig3} 
\end{figure*}

\textit{Time-dependent environments:} 
Motivated by periodic treatments, as used in microbial and cancer cells, we consider repeated on-off-switching of the death process (Fig.~\ref{fig1}, bottom right). We let $\de(x,t)=\de(x)$ in an on-environment of length $T_\text{on}$ and $\de(x,t)=0$ in an off-environment of length $T_\text{off}$ while cells keep dividing regardless $\g(x,t)=\g(x)$. Stochastic simulations display a complex landscape of survival probabilities with several peaks, Fig.~\ref{fig3}a,b.

We consider the offspring distribution of the underlying branching process to calculate the survival probability. Here, the generating functionals of the offspring distributions originating from a cell with age $x_0$ after an on- or off-environment starting at time $0$ and ending at $T_\text{on/off}$ are $\Zon [h,T_\text{on}|\delta_{x_0}]:=\int \mathcal{D}[\n] \exp(\int_0^{\infty}{\rm{d}}x\ln(h(x))\n(x))\prfun[\n,T_\text{on}|\delta_{x_0}]$ and similarly $\Zoff[h,T_\text{off}|\delta_{x_0}]$. Note that the auxiliary variable is a general age-dependent function $h(x)$, \changes{which is dimensionless}, and the integral is over non-negative measures $n$ with positive support. The generating functional of offspring after each period is then $\Zon[\Zoff[h,T_\text{off}|\delt],T_\text{on}|\delta_{x_0}]$, where $\bullet$ is a placeholder for the function argument. Since the environments are periodic, offspring distributions are invariant across environments after one period. This means that the survival probability $\psurv(x_0)$, starting from a single cell with age $x_0$, is obtained from the embedded Galton-Watson chain that satisfies:
\begin{equation}\label{fixed}
1-\psurv(x_0)=\Zon[\Zoff[1-\psurv,T_\text{on}|\delt],T_\text{off}|\delta_{x_0}]
\end{equation}
as a functional generalisation of the multi-type branching process \cite[Ch. II.7]{Harris1964}. 

We now derive a method to calculate the generating functional of the offspring distributions from the Master equation \eqref{eq: master_equation}. The standard solution method \cite{gardiner1985handbook} involves transforming the Master equation into a partial differential equation (PDE) for the probability generating function and solving it using the method of characteristics. The method can be adapted to the age-structured case using the generating functionals of offspring after an on- or off-environment of length $T_\text{on/off}$. 
To this end, we transform the functional Master equation~\eqref{eq: master_equation} into a functional differential equation for $\Zon$:
\begin{align}\label{Zevolution}
      &\frac{\partial \Zon[h,t|\delta_{x_0}]}{\partial t}=\int_0^{\infty}\!\!{\rm{d}}x\Big[\partial_xh(x)-(\g(x)+\de(x))h(x)\notag\\
      &\qquad\qquad  +\g(x)h^{2}(0)+\de(x)\Big]\frac{\delta \Zon[h,t|\delta_{x_0}]}{\delta h(x)} .
\end{align}
The characteristics of Eq.~\eqref{Zevolution} obey PDEs, whose solution satisfies a non-linear integral equation (\App~\ref{app:characteristics}):
\begin{align}\label{BH}
&\Zon[h,t|\delta_{x_0}]\Pi(x_0)=\Pi(t+x_0)h(t+x_0)+\int_0^t\!\!{\rm{d}}u\,\Pi_d(u+x_0)\notag\\
&\qquad\qquad +\int_0^{t}{\rm{d}}u\,\Pi_{\gamma}(u+x_0)\Zon^2[h,t-u|\delta_{0}],
\end{align}
and we simply set $d(x)=0$ in an off-environment to obtain $\Zoff$. The function $\Pii(x)=e^{-\int_0^x{\rm d}u(\g(u)+\de(u))}$ is the probability for a newborn cell to survive until age $x$ and $\Pig(x)= \gamma(x)\Pii(x)$ and $\Pid(x)=d(x)\Pii(x)$ are the probabilities for a newborn cell to divide or die at age $x$, respectively. Eq.~\eqref{BH} generalises the integral equation derived by Bellman and Harris \cite{Harris1964} similar to \cite{sevast1964age,Kimmel1983} where cell death occurs independently from divisions, but here we account explicitly for division-/death rates. Using Eq.~\eqref{BH} in \eqref{fixed} gives a system of integral equations that can be solved using an iterative numerical procedure (\App~\ref{TD-environments}). 

In agreement with simulations, the survival probability displays narrow peaks of survival probabilities that emerge when \changes{the environment period $P=T_\text{on}+T_\text{off}$ is tuned precisely to the mean division $\mg$}. These peaks are absent in the unstructured models with constant division and death rates (\App~\ref{Unst-TD-environments}). Reminiscent of overtones, the peaks repeat for higher integer multiples of $P$, and we therefore call this phenomenon {\it survival resonances}.  

We find two types of resonances. Firstly, \textit{deterministic resonances} occur when division and death times fluctuate little, i.e.~$\cvd,\cvg\ll1$, Fig.~\ref{fig3}a. In this limit, division and death occur at fixed ages $\mg$ and $\md$. If $\mg<\md$, death never occurs because no cell ever reaches the age of death $\md$, and $\psurv(0)=1$, see Fig.~\ref{fig3}c. If $\mg>\md$, cells die if they reach the age of death $\md$ in any of the on-environments \changes{$A_m:=mP+[0,T_\text{on}]$}. 
Descendants synchronously exceed the deadly age $\md$ in time intervals $B_n=n\mg+(\md,\mg], n\in\mathbb{N}_0$. Moreover, all of them die simultaneously if this occurs in any on-environment $A_m$, leading to the survival condition that $\forall m,n: A_m\cap B_n=\varnothing$ implies $\psurv(0)=1$. Consequently, only singular, fine-tuned combinations of $T_\text{on,off}$ and $\mg$ allow  deterministic resonance:\changes{\begin{align}\label{det_res}
    &\{T_\text{on,off}:\mg=nP,n\in\mathbb{N},0<\mg-\md<T_\text{off}\}.
\end{align} }
Physically, Eq.~\eqref{det_res} is a renewal condition: all the generations are approximately born at the beginning of an on-environment.

We make the following observations on the effect of noise seen in Fig.~\ref{fig3}a. Firstly, higher overtones of survival probabilities are progressively attenuated because they have a higher absolute variance of division ages and therefore are less tuned to \changes{$P$}. Secondly, the asymmetry and shift of peaks are due to noise since division times longer than \changes{$n P$} lead to death, while cells with shorter division times avoid it.
As more and more noise in division timing is introduced (Fig.~\ref{fig3}a), deterministic resonances disappear and give way to the second type of resonance, which we call \textit{noise-induced resonances}. Here, peaks in $\psurv$ appear for specific combinations of $\md$ and \changes{$T_\text{on,off}$} with only small fluctuations in death times. These resonances also appear for $\mg < \md$, and their emergence is qualitatively explained by considering the death probability of the first cell. Assuming that death is deterministic \changes{($\cvd\rightarrow0$) with putative death time $\tau_d\approx \md$}  and that the cell was born at time $t=0$ at the start of the first on-environment, we can identify two regimes. If $\tau_d$ falls in an on-environment, the death probability is the product of the probability $w_d$ that death would occur at time $\tau_d$ times the probability that the cell does not divide before $\tau_d$, $\phi_\gamma(\tau_d)=e^{-\int_0^{\tau_d}\rm{d} u \gamma(u) }$. If $\tau_d$ falls in an off-environment, the cell either dies at the next on-environment \changes{($w_d=1$) or never ($w_d<1$) due the zero death rate at old ages, see Fig.~\ref{fig2}b}.  This consideration leads to the effective death and persistence probability of the first cell:
\changes{
\begin{align}
	\nu_d \approx& \mathbb{E}
	\left[ \mathbbm{1}_\text{on}(\tau_d)w_d\phi_\gamma(\tau_d)+\mathbbm{1}_\text{off}(\tau_d)\delta_{w_d,1}\phi_\gamma\left(\left\lfloor \frac{\tau_d}{P}+1\right\rfloor P\right) \right]\notag\\
 \nu_\varnothing\approx&(1-w_\gamma)\mathbb{E}\left[\mathbbm{1}_\text{on}(\tau_d)(1-w_d)+\mathbbm{1}_\text{off}(\tau_d)(1-\delta_{w_d,1})\right],\label{eff_death_1st}
\end{align}
where $\delta_{w_d,1}$ is the Kronecker-$\delta$}, $\mathbbm{1}_\text{off/on}(\tau_d)$ is $1$ if $\tau_d$ falls into an off/on environment and zero otherwise and the average is over the distribution of death times $\tau_d$. \changes{Eq.~\eqref{eff_death_1st} can also be derived from Eq.~\eqref{BH} by neglecting the last term, which does not contribute to the first generation. The formula explains how death events are filtered through cell division and the environment. }
As a rough approximation, we assume that the cells of later generations have the same death probability as the first cell and then use Eq.~\eqref{eff_death_1st} in Eq.~\eqref{pext}.  These approximations, strictly valid only for constant environments, predict resonances at the following environment lengths:
\changes{\begin{align}\label{eq_noise-induced_resonance}
	\{T_\text{on,off}:\md=T_\text{on}+nP, n\in\cchanges{\mathbb{N}_0}\},
\end{align} 
which} reproduce all the noise-induced resonances seen in the simulations and numerics (dashed lines, Fig.~\ref{fig3}a,b). \changes{Intuitively, resonances occur when a cell's putative death time falls shortly after the end of an on-environment because this maximises its time to divide.} The approximation produces ramps followed by sudden drops and thus provides qualitative agreement with the exact results. We also calculated the transient survival probabilities, which show that the first generation determines the locations of all noise-induced resonances in the following generations (\App~\ref{appendix_noise-induced_resonances}). \changes{Neither changes in $\frac{T_\text{on}}{T_\text{off}}$ nor the sign of the growth rate $\lambda$ in the on-environment affect the existence and locations of noise-induced resonances. Such changes only affect their amplitudes (SM Fig.~\ref{sfig:robustness}). Resonances were significantly boosted in cases with persistence ($\nu_\varnothing>0$, Fig.~\ref{fig3}b and SM Fig.~\ref{sfig:robustness}b). This is explained by resonant filtering through a decreasing death rate for old cells (Fig.~\ref{fig2}b inset, Eq.~\eqref{eff_death_1st}). Predictions of the resonance profile, including the ramp shape and finite survival probability, improved when transients beyond the approximation in Eq.~\eqref{eff_death_1st} were taken into account (Fig.~\ref{fig3}b red dotted-dashed line, SM~\ref{appendix_noise-induced_resonances}).} We summarise the locations of deterministic and noise-induced survival resonances in a phase diagram (Fig.~\ref{fig3}c), which is independent of persistence.  

\textit{Discussion:} We examined a stochastic population model with cell cycle-dependent division and death rates to understand how fractional killing may emerge from the interplay of cellular heterogeneity with drug treatments. 
We found that survival chances increase with noise in division timing but decrease with noise in death timing.   This suggests that division heterogeneity combined with controlled cell death could represent a strategic advantage for cells to persist in continuous drug treatment.

Then, we analysed survival probabilities under periodic treatments and discovered deterministic and noise-induced resonances observed under a tuning of division and death rates to the environment duration. \changes{Doubling times of cancer cell lines vary from one to several days but lengthen significantly in tumours \cite{mitchison2012proliferation} and after drug exposure \cite{mohammadi2022lineage}. In comparison, treatment lengths are experimentally (one to several days in cell lines) and clinically set parameters (months), and how to tune these optimally is subject to ongoing research \cite{maltas2024drug}. Survival resonances should thus be observable within experimental and clinical parameter ranges.} 

In populations with pre-existing genetic variability, such as in genetic screens or cancer, these resonances could be difficult to avoid as division and death rates are heterogeneous and can be selected by adjusting the environmental duration. The survivors of this selection process will be located predominantly along peaks of the survival probability, shown as lines in the phase diagram, Fig~\ref{fig3}c.
Under this hypothesis, deterministic resonances select low-noise phenotypes within a narrow band of division times that are even integer multiples of the environmental duration, Eq.~\eqref{det_res}.  
Noise-induced resonances select on cells with death times being odd multiples of the environmental duration. 

Interestingly, for $\mu_d > \mu_\gamma$, the resonance increases with division time noise and, in some cases, reaches a maximum (SM Fig.~\ref{stochastic_res}), which is a hallmark of stochastic resonance \cite{gammaitoni1998stochastic}. The intuition is that cells are killed when they stochastically exceed an age-dependent death threshold and these events are resonantly filtered by periodic forcing.
Threshold fluctuations attenuate resonances but enhance the survival probability away from them (Fig.~\ref{fig3}b) in contrast to constant environments. Complex intracellular pathways implement such thresholds \cite{Spencer2009,roux2015fractional,Ballweg2017} and our findings could link their noise properties with evolvable survival strategies that can be exploited for decision-making or synthetic biology. 

A limitation of our approach is that we ignored generational inheritance \cite{nozoe2020cell,hughes2022patterns}, \changes{for example, through cell size control in bacteria. This can be included by averaging Eq.~\eqref{eff_death_1st} over the size distribution, which would deviate from the ramp-like behaviour of survival resonances but not their location. Another limitation is that we neglected effects that are dependent on drug exposure time \cite{wakamoto2013dynamic},} which could be incorporated by generalising Eq.~\eqref{BH} to the time-dependent case (SM \ref{Characteristic_curves_Master_Equation}). %It would be interesting to explore extensions to random environments with fluctuating drug duration. Such random environments likely select several resonances along a vertical section of the phase diagram (Fig.~\ref{fig3}(c)) but these effects are beyond the scope of our study.

In summary, we have identified a possible non-genetic mechanism for fractional killing emerging in response to periodic drug treatments. These stochastic mechanisms could be one of many contributing factors to drug persistence, among other active mechanisms such as phenotypic switching \cite{kobayashi2015fluctuation}. Since periodically forced age-structured branching processes are ubiquitous, such as in epidemics and ecology, we anticipate survival resonances to be widespread in stochastic population dynamics. 

\textit{Acknowledgments:} We thank Farshid Jafarpour for insightful comments on the manuscript. \textit{Funding:} This work was supported through a UKRI Future Leaders Fellowship (MR/T018429/1 to PT) and a EPSRC DTP studentship (EP/T51780X/1 to FP).

\bibliography{bibliography}
\newpage
\onecolumngrid
\setcounter{figure}{0}
\renewcommand{\thefigure}{S\arabic{figure}}

\renewcommand\appendixname{Supplemental Material}
\appendix
\newpage
\def\Ton{T_\text{on}}
\def\Toff{T_\text{off}}

\begin{center}
	\Large\textbf{Supplemental materials}
\end{center}

\setcounter{secnumdepth}{2}
\section{Offspring generating functional in time-dependent environments}\label{Characteristic_curves_Master_Equation} 
\label{app:characteristics}
Here, we provide the general integral equations characterising the offspring distributions in a time-dependent environment, which contains~\eqref{BH} as a special case. The probability $\prfun[n,t|\delta_{x_0},t_0]$ of observing a population with number density $n(x)$ at time $t$ arising from a single ancestor of age $x_0$ at time $t_0$. The probability generating functional associated with this offspring distribution is given by 
$\genfun [h,t|\delta_{x_0},t_0]:=\int \mathcal{D}[\n] \exp(\int_0^{\infty}{\rm{d}}x\ln(h(x))\n(x))\prfun[\n,t|\delta_{x_0},t_0]$ where the auxiliary variable is a general age-dependent function $h(x)$ and the integral is over all non-negative measures $n$ with positive support. Using the definition of the generating functional in the Master equation~\eqref{eq: master_equation} we find a functional derivative equation:
\begin{equation}\label{app_eq::Zpde}
	\begin{split}
		\frac{\partial\genfun[h,t|\delta_{x_0},t_0]}{\partial t}=&\int_0^{\infty}\big[\partial_xh(x)-(\g(x,t)+\de(x,t))h(x)+\g(x,t)h^2(0)+\de(x,t)\big]\frac{\delta \genfun[h,t|\delta_{x_0},t_0]}{\delta h(x)}{\rm{d}} x, \\
	\end{split}
\end{equation}
which can be manipulated via the method of characteristics. Doing so, the characteristic curves are parameterised in terms of a continuous variable $s$ and are described by two ODEs:
\begin{equation}\label{app_eq::odes}
	\frac{dt(s)}{ds}=1,\ \
	\frac{d\genfun(s)}{ds}=0 
\end{equation}
and a PDE for \(h(x,s)\): 
\begin{equation}\label{app_eq::h_pde}
	\begin{split}
		& (\partial_s+\partial_x)h(x,s)=(\de(x,s)+\g (x,s))h(x,s)-\g(x,s)h(0,s)^2-\de(x,s).\\
	\end{split}
\end{equation}
We define $\Pi(s,x)=\exp\Bigl(-\int_0^s\g(x+u,u)+\de(x+u,u){\textrm{d}}u\Bigr)$ and integrate Eq.~\eqref{app_eq::h_pde} to obtain:
\begin{equation}\label{app_eq::h_int}
	h(x,0)=\Pi(s,x)h(x+s,s)+\int_0^{s}\Pi(y,x)\big(\g(y+x,y)h^2(0,y)+\de(y+x,y)\big)dy.
\end{equation}
We fix $t_0=0$ so that $t(s=0)=0$ and $\genfun[h(\bullet,0),t(0)|\delta_{x_0},0]=h(x_0,0)$. Under these initial conditions, we obtain $t(s)=s$ and $\genfun[h(\bullet,s),t(s)|\delta_{x_0},0]=h(x_0,0)$ from  Eqs.~\eqref{app_eq::odes}. Therefore, Eq.~\eqref{app_eq::h_int} can be expressed as:
\begin{equation}\label{app_eq::Z_int}
	\genfun[h(\bullet,s),s|\delta_{x_0},0]=\Pi(s,x)h(x+s,s)+\int_0^{s}\Pi(y,x)\big(\g(y+x,y)h^2(0,y)+\de(y+x,y)\big)dy.
\end{equation}
The final step is replacing $h^2(0,y)$ with a known function in Eq.~\eqref{app_eq::Z_int}. It follows from Eq.~\eqref{app_eq::h_pde} that $\mathcal{Z}$ is constant along any characteristic starting at $s=y$ and hence:
\begin{equation}\label{app_eq::ic_shift}
	h(x_0,y)=\genfun[h(\bullet,s),t(s)|\delta_{x_0},y].
\end{equation}
Plugging Eq.~\eqref{app_eq::ic_shift}
in Eq.~\eqref{app_eq::Z_int}, we obtain an integral equation for the offspring initial generating function in time-dependent environments:
\begin{equation}\label{app_eq::BH-timedependent}
	\genfun[h,t|\delta_{x_0},0]=\Pi(t,x_0)h(x_0+t)+\int_0^{t}\Pi(y,x_0)\bigg(\g(x_0+y,y)\genfun^2[h,t|\delta_{0},y]+\de(y+x_0,y)\bigg)dy.
\end{equation}

Finally, we restrict our analysis to constant environments (so that $\genfun[h,t+T|m,T]=\genfun[h,t|m,0]$ for any range of time $T$) and adopt the notation proposed in Eq.~\eqref{BH}:
\begin{equation}\label{app_eq::BH}
	\begin{split}
		\genfun[h,t|\delta_{x_0},0]\Pi(x_0) =\Pi(t+x_0)h(t+x_0)+\int_0^{t}\Pi_{\de}(u+x_0)du+\int_0^t\Pi_{\g}(u+x_0)\genfun^2[h,t-u|\delta_0,0]du\\
	\end{split}
\end{equation}
The long-term extinction probability follows from $\lim_{t\rightarrow\infty}\genfun[h=0,t|\delta_{x_0},0]=p^*(x_0)=1-\psurv(x_0)$ in Eq.~\eqref{app_eq::BH}, which gives
\begin{equation}\label{app_eq::pext}
	p^*(x_0)\Pi(x_0)=\int_{x_0}^{\infty}\Pi_{\de}(u)du+(p^*(0))^2\int_{x_0}^{\infty}\Pi_{\g}(u)du.\\
\end{equation}
Setting $x_0=0$ and identifying $\nu_{\de}=\int_0^{\infty}{\rm{d}}x\Pi_{\de}(x)$ and $\nu_{\g}=\int_0^{\infty}{\rm{d}}x\Pi_{\g}(x)$ as the effective death- and division probabilities, respectively, Eq.~\eqref{pext} of the main text follows.
Together with Eq.~\eqref{app_eq::pext}, this provides an exact solution for the extinction probability of a population starting from a single ancestor with initial age~$x_0$.

\section{Long-term survival probability in periodic on-off environments}\label{TD-environments}
Here, we present a detailed derivation the equations governing the survival probability during periodic on-off treatments.  such framework which, in substance, is an age-dependent generalization of the multi-type Galton-Watson processes~\cite[Ch. II.7]{Harris1964}.
We start observing that, for any branching processes, the evolution of each branch is independent. This means that the generating functional for a number of branches with density $q(x_0)$ can be expressed in terms of evolution of each individual branch starting with age $x_0$: 
\begin{equation}\label{app_eq::branching_property}
	\genfun[h,t|q,t']= \exp\int_0^{\infty} q(x_0)\ln(\genfun[h,t|\delta_{x_0},t'])dx_0,
\end{equation}
for $t'< t$. Using the above in the Chapman-Kolmogorov integral equation, $\genfun[h,t|m,0]=\int \genfun[h,t|q,t']\prfun[q,t '|m,0]\mathcal{D}q$, gives:
\begin{align}\label{app_eq::chapman-kolmogorov-branching}
	\genfun[h,t|m,0]&=\int e^{-\int_0^{\infty}q(x)\ln(\genfun[h,t|\delta_x,t']dx)}\prfun[q,t '|m,0]\mathcal{D}q.
\end{align}
Eq.~\eqref{app_eq::chapman-kolmogorov-branching} can be expressed via nesting generating functionals:
\begin{align}
	\label{app_eq::embedding_z}
	\genfun[h,t|m,0]&=\genfun[\genfun[h,t|\delta_{\tiny{\bullet}},t'],t'|m,0],
\end{align}
where $\bullet$ is a placeholder for the function argument, i.e.,  $\genfun[h,t|\delt,t'](x_0)=\genfun[h,t|\delta_{x_0},t']$. Furthermore, we model periodic on-off treatment by nesting the generating functionals of on-environment $\Zon[h,T_\text{on}|\delta_{x_0},0]$ and off-environments $\Zoff[h,T_\text{off}|\delta_{x_0},0]$.
Since the on-off environment sequence is repeated with period $P=\Ton+\Toff$, we define the offspring generating functional at time $t=P$ as:
\begin{equation}\label{app_eq::offspring_generating_W}
	\mathcal{W}[h,x_0]=\mathcal{W}[h,\bullet]({x_0})=\Zon[\Zoff[h,\Toff|\delt,0],\Ton|\delta_{{x_0}},0].
\end{equation}
Therefore, the offspring generating functional evaluated at time $t=kP$ (with $k \in \mathbb{N}$) can be expressed by nesting Eq.~\eqref{app_eq::offspring_generating_W} $k$ times:
\begin{equation}
	\genfun[h,t=kP|\delta_{x_0},0]=\mathcal{W}^{k}[ h,x_0],
\end{equation}
where we defined the  $k^{th}$ composition of functions as $\mathcal{W}^k[h;x_0]=(\mathcal{W}\circ\mathcal{W}^{k-1})[h,x_0]=\mathcal{W}[\mathcal{W}^{k-1}[h,\tiny{\bullet}];x_0]$.\\
Following a similar argument to Harris~\cite{Harris1964},  we set $h=0$ to obtain  the extinction probability, for which we assume the existence of an asymptotic steady state for $t>kP$: 
\begin{equation}     \mathcal{W}[\mathcal{W}^{k}[0,\bullet],x_0]=
	\mathcal{W}[0,x_0]. 
\end{equation}
Finally, we obtain an  expression (equal to Eq.~\eqref{fixed}) for the asymptotic extinction probability  $p^*(x_0)=1-\psurv(x_0)$ of a population starting from a single individual with age $x_0$:
\begin{equation}\label{app_eq::extinction_fixed_point}
	\begin{split}
		p^*(x_0)=\mathcal{W}[p^*,x_0]=\Zon[\Zoff[p^*,\Toff|\delt,0],\Ton|\delta_{x_0},0].
	\end{split}
\end{equation}
Manipulating Eq.~\eqref{app_eq::extinction_fixed_point}~via~Eq.~\eqref{BH}, we obtain a  system of integral equations:
\begin{equation}\label{app_eq::eq_iterative_integral_equationsA}
	\begin{split}
		\mathcal{G}[x_0,t;p^*]\Pi^\text{on}(x_0)&=\Pi^\text{on}(x_0+t)\mathcal{H}[x_0+t,t;p^*]+\int_0^t\Pi^\text{on}_{\gamma}(u+x_0)\big( \mathcal{G}[0,t-u;p^*]\big)^2du+\int_0^t\Pi^\text{on}_d(u+x_0)du,
	\end{split}
\end{equation}

\begin{equation}\label{app_eq::eq_iterative_integral_equationsB}
	\begin{split}
		\mathcal{H}[x_0,t;p^*]\Pi^\text{off}(x_0)&=\Pi^\text{off}(x_0+t)\mathcal{G}[x_0+t,\Toff;p^*]+\int_0^t\Pi^\text{off}_{\g}(x_0+u)(\mathcal{H}[0,t-u;p^*])^2du,
	\end{split}
\end{equation}
where we defined
$\mathcal{G}[x,t;p^*]=\Zon[\Zoff[p^*,t|\delt,0],\Ton|\delta_x,0]$  and 
$\mathcal{H}[x,t;p^*]=\Zoff[p^*,t|\delta_{x_0}]$.
Eqs.~\eqref{app_eq::eq_iterative_integral_equationsA} and \eqref{app_eq::eq_iterative_integral_equationsB} represent a system of coupled integral equation that can be solved for $p^*(x_0)=\mathcal{G}[x_0,\Toff;p^*]$.
Moreover, the probabilities with superscript $^\text{on}$ and $^\text{off}$ are the defined as in Eq.~\eqref{BH} with rate functions respectively $\{\g(x),\de(x)\}$  and $\{\g(x),0\}$. Eqs.~\eqref{app_eq::eq_iterative_integral_equationsA} and~\eqref{app_eq::eq_iterative_integral_equationsB} can be solved numerically by 1) discretising functions and integrals and 2) iteratively inserting them into each other until the fixed point is reached.

\section{Constant rates in time-dependent environments }\label{Unst-TD-environments}
We discuss the age-independent branching processes in a periodic on-off environment. Constant rates imply exponentially distributed division and death times with $w_d=w_\gamma=1$. For simplicity, we assume $T_\text{on}=T_\text{off}=T$. 

In a constant environment, an exact solution of the generating function can be obtained~\cite{Kendall1948b}:
\begin{equation}\label{app_eq::unstructured_gen_fun}
	\genfun(h,T|1,0)=\frac{\de+\de(h-1)e^{(\g-\de)T}-\g h}{\de+\g(h-1)e^{(\g-\de)T}-\g h},
\end{equation}
where $\de$ and $\g$ are the age-independent death and division rates (respectively equal to the inverse of the average death time $\mu_{\de}$ and average division time $\mu_{\g}$) and $h$ is the age-independent auxiliary variable. The condition $|1,0$ represents the initialisation of the system with one individual at time $t=0$.
From comparison with Eq.~\eqref{pext}, we find the survival probability from Eq.~\eqref{app_eq::unstructured_gen_fun}: $\psurv=1-\lim_{T\rightarrow\infty }\genfun(h=0,T|1,0)=
\frac{\g-\de}{\g}\Theta(\frac{\g-\de}{\g})$, where $\Theta$ is the Heaviside function.

\begin{figure}[h!]
	\centering
	\includegraphics[width=5cm]{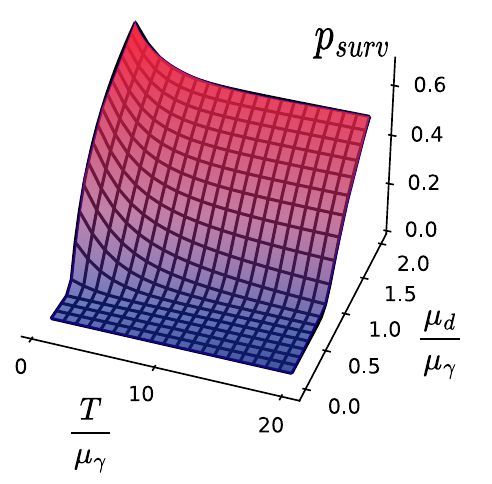}
	\caption{\textbf{Survival resonances are absent for constant division and death rates.} Asymptotic, age-independent survival probability in a periodic time-dependent environment, Eq.~\eqref{app_eq::unstructured_extinction_time_dependent}. }
	\label{fig:unstru}
\end{figure}

We now move to time-dependent environment characterised by a fixed division rate $\g$  and a periodically switching death rate $\de(t)$. The asymptotic extinction probability function $\psurv$ solves the following fixed point equation: $1-\psurv=\Zon(\Zoff(1-\psurv,T|1,0),T|1,0)$, where $\Zon$ and $\Zoff$ labels the generating function for on- and off-environments. This gives:
\begin{equation}\label{app_eq::unstructured_extinction_time_dependent}\psurv=1-\frac{\de\big(e^{2\g T}-e^{(\de+\g)T}\big)}{\de e^{\de T}-\de e^{(\de+\g)T}-\g\big(e^{\de T}+e^{2\g T}\big)}.
\end{equation}
As shown in Fig.~\ref{fig:unstru}, we observe a phase transition between extinction (blue, $\psurv=0$) and fractional killing (red shading, $\psurv>0$). In contrast to age-dependent results, the survival probability is monotonic in $\frac{\md}{\mg}$ and $\frac{T}{\mg}$ and there are no survival resonances. 

\section{Alternative approximation through transient behaviour of the survival probability}\label{appendix_noise-induced_resonances}

The extinction probability for the first generation (i.e.~the first cell born at $t=0$) was given in Eq.~\eqref{eff_death_1st} by:
\begin{equation}\label{1-st}
	p^{I}_\text{ext}(\md) =  \mathbbm{1}_\text{on}(\md)w_d\phi_\gamma(\md)+\mathbbm{1}_\text{off}(\md)\delta_{w_d,1}\phi_\gamma\left(\left\lfloor \frac{\md}{P}+1\right\rfloor P\right),
\end{equation}
We observe that in the second generation, the two newborns have a synchronised age and thus their extinction probability $p^{II}_\text{ext}(\md,\tau_d)$ for a putative death time $\tau_d$ equals:
\begin{equation}\label{2-st}
	\begin{split}
		p_\text{ext}^{II}(\md,\tau_d)&=\int_0^{\tau_d}p(\tau_{\gamma}=x|\tau_{\gamma}<\tau_d)(p^{I}_\text{ext}(\md+x))^2dx=\\
		&=\int_0^{\tau_d}\frac{f_\gamma(x)}{\int_{0}^{\tau_d}f_\gamma(u){\rm d}u}(p^{I}_\text{ext}(\md+x))^2dx\\
	\end{split}
\end{equation}
If $w_d=1$, then $\tau_d=\mathbbm{1}_\text{on}(\md)\md+\mathbbm{1}_\text{off}(\md)\lfloor \frac{\md}{P}+1\rfloor P$. If $w_d<1$, then $\tau_d=\mathbbm{1}_\text{on}(\md)\md+\mathbbm{1}_\text{off}(\md)\infty$. This allows reducing $p_\text{ext}^{II}(\md,\tau_d)$ to $p_\text{ext}^{II}(\md)$.

\begin{figure*}[h!]
	\includegraphics[width=16cm]{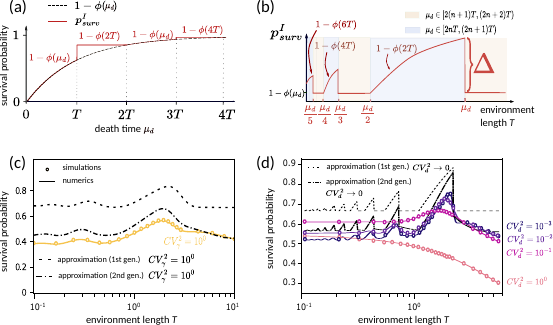}
	\caption{\textbf{Transient behaviour of the survival probability in periodic on-off environments.} $w_d=w_\gamma=1\Rightarrow\nu_\emptyset=0$. (a) Survival probability of the initial cell ($\psurv^{I}$, first-generation approximation) as a function of $\md$ (red line) is discontinuous at multiples of the environment length $T=T_\text{on}=T_\text{off}$. $1-\phi(\mu_d)=1-e^{-\int_0^{\mu_d}{\rm d}u\gamma(u)}$ is the corresponding survival probability in a constant environment, assuming $CV^2_d\rightarrow 0$. (b) Noise-induced jumps of the survival probability in the first generation ($\psurv^{I}$) as a function of the environment length $T$. The height $\Delta$ of the first resonance peak is analytically approximated via Eq.~\eqref{jumps}, see also Fig.~\ref{stochastic_res}. (c) The survival probabilities of the first and second generations (Eqs.~\eqref{1-st}~and~\eqref{2-st}) approximate the asymptotic behaviour (yellow line). (d) First- and second-generation approximations (Eqs.~\eqref{1-st}~and~\eqref{2-st}) of noise-induced resonances, as shown in Fig.~\ref{fig3} with $\md=2.0$, $\mg=1.8$ and $\cvg=1$. 
		Non-monotonic behaviour is also observed for finite noise in death time ($\cvd\in\{10^{-1},10^{-2},10^{-3}\}$) except for the age-independent case ($\cvd=1$, rose line). Parameters are $\md=1.8$, $\cvd=10^{-2}$, $\mg=2.0$ and $\cvg=1$ as in Fig.~\ref{fig3}~(b).}  \label{Four_figs}
\end{figure*}

The probability that the first cell divided equals $p^{I}_\text{div}=1-p^{I}_\text{ext}$ in the case without persistence, and in the case with persistence\begin{equation}
	p^{I}_\text{div}(\md)=\mathbbm{1}_\text{on}(\md)w_\gamma\left(w_d\int_{0}^{\md}f_\gamma(u){\rm d}u+(1-w_d)\right)+\mathbbm{1}_\text{off}(\md)w_\gamma
\end{equation}
\begin{equation}\label{eq:2nd_gen_approx}
	p_{\text{ext}\le2}(\md)=p^{I}_\text{ext}(\md)+p^{I}_\text{div}(\md)\cdot p^{II}_\text{ext}(\md)
\end{equation}
The above equation derives from Eq.~\eqref{BH}. In Fig.~\ref{Four_figs} (c, d), we compare the survival probability of the first and second generations and how they approach the asymptotic survival probability. The effect of persistence is shown in Fig.~\ref{fig3}b.

\subsection*{Stochastic resonance behaviour of noise-induced resonances}

\begin{figure*}[h!]
	\includegraphics[width=7.5cm]{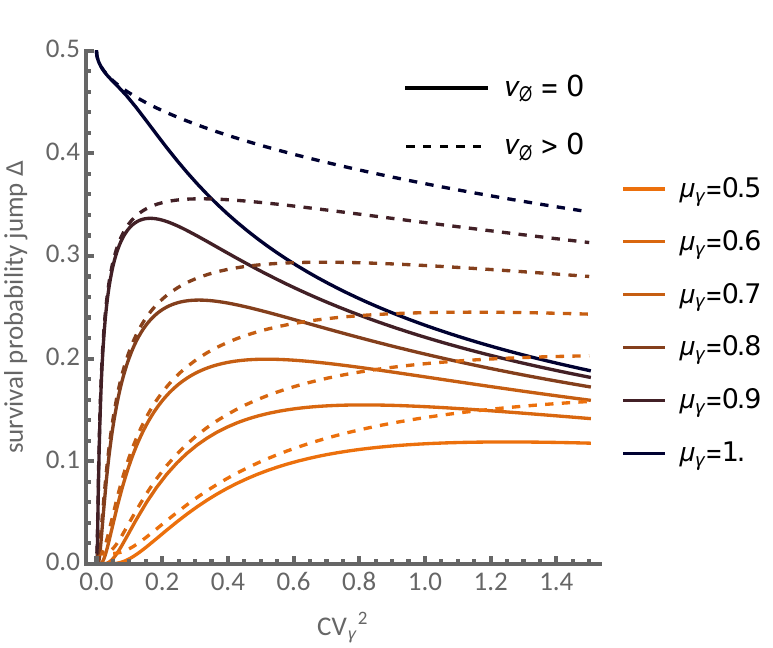}
	\caption{\textbf{Stochastic resonance-like behaviour of the survival probability peaks.} Dominant jump (of the survival probability in the first generation) versus the noise in division times at $\frac{\md}{T_\text{on}}=1$. For some values of $\mg$, the survival jumps display a non-monotonic behaviour following Eq.~\eqref{jumps} (solid lines, $w_d=w_\gamma=1$). For the case with persistence (dashed lines, $w_d=w_\gamma=0.99$), the peaks broaden displaying an amplification of the survival probability over a wide range of noise levels in division times. }  \label{stochastic_res}
\end{figure*}

We conclude by discussing the relation of noise-induced survival resonances and stochastic resonance. To this end, we study the relationship between noise-induced jumps in $\psurv$ and the noise affecting division times. The magnitude $\Delta$ of the first resonance peak of the survival probability can be approximated by the corresponding magnitude in the survival probability of the first generation. Based on Eq.~\eqref{1-st}, we find:
\begin{equation}\label{jumps}
	\Delta\approx w_d\phi_\gamma(T_\text{on})-\delta_{w_d,1}\phi_\gamma\left(\left\lfloor \frac{T_\text{on}}{P}+1\right\rfloor P\right)%\phi(2T)-\phi(\md)=e^{-\int_0^{\md}\gamma(u)du}\big(1-e^{-\int_{\md}^{2T}\gamma(u)du}\big).
\end{equation}
In Fig.~\ref{stochastic_res} we use this equation to quantify the output-performance ($\Delta$) versus the input noise magnitude ($CV^2_{\g}$) for $\Gamma$-distributed division times. We observe that, for specific values of $\md$, the jump height displays a non-monotonic dependence on $\cvg$. The peaks indicate an increase in the signal-to-noise ratio with the noise intensity, which is the signature of stochastic resonance.

{
	\section{Summary of mean-field results}
	\label{SM::growthrate}
	
	\def\mf{\overline{n}}
	\def\mfp{\mf_+}
	\def\mfm{\mf_-}
	\def\mfpm{\mf_\pm}
	
	Here, we recapitulate the mean-field results in a constant environment, including a non-zero persister fraction within the long-time limit. The mean-field density $\mf(x,t)= \int \mathcal{D}[n] n(x) \mathcal{P}[n,t|\delta_0]$ obeys the McKendrick-von-Foerster equation:
	\begin{align}
		&\left(\partial_t +\partial_x +\gamma(x)+d(x)\right) \mf(x,t) = 0,\ \
		\mf(0,t) =2 \int_0^\infty {\rm d} x \gamma(x) \mf(x,t).
	\end{align}
	The above follows from multiplying the master equation \eqref{eq: master_equation} by $n(x)$ and integrating.
	Setting $\Pi(x) = \exp\bigl(-\int_0^x {\rm d}s (\gamma(s) +d(s))\bigr) $ as before, we obtain the solution:
	\begin{align}
		\mf(x,t) = \Theta(t-x)\Pi(x) \mf(0,t-x) + \delta(x-t) \Pi(x),
	\end{align}
	where the first term stems from newborn cells and the second term from the initial cell. The newborn density has the asymptotic form $\mf(0,t) \sim \eta_0 e^{\lambda t}$ and $\eta_0$ is a constant that depends on the initial condition. Here $\lambda$ is the real solution of the Euler-Lotka equation
	$1=2 \int_0^\infty {\rm d} x e^{-\lambda x}\Pi(x)  \gamma(x)$, which quantifies  the exponential growth rate of newborn cells $\mf(0,t) \sim \eta_0 e^{\lambda t} $.
	
	We now decompose the solution $\mf(x,t)=\mfp(x,t)+\mfm(x,t)$ into normally proliferating and persisting cells, respectively. Then since $\Pi(\infty)=\nu_\varnothing$ is the persistence probability, we identify $c^{-1}(\Pi(x)-\Pi(\infty))e^{-\lambda x}$ as the stable age-distribution of the proliferating cells, where $c$ is a normalising constant, and write:
	\begin{align}
		&\mfp(x,t) \sim \Theta(t-x)\eta_0 e^{\lambda(t-x)} \left(\Pi(x)-\Pi(\infty)\right)+\delta(x-t)\left(\Pi(x)-\Pi(\infty)\right),\notag\\
		&\mfm(x,t) \sim  \Theta(t-x) \eta_0 e^{\lambda(t-x)}\Pi(\infty) + \delta(x-t) \Pi(\infty).
	\end{align}
	We obtain total abundances by integrating these over cell age $x$, $N_\pm (t)= \int_0^\infty {\rm d}x \mfpm(x,t)$,  gives
	\begin{align}
		N_{+}(t)  \sim  \eta_0 e^{\lambda t} c+\Pi(t)-\Pi(\infty),\ \ 
		N_{-}(t) \sim  \Pi(\infty) \left(1 + \eta_0 \frac{(e^{\lambda t} -1)}{\lambda}\right).
	\end{align}
	Thus, on average, if $\lambda>0$ both the persisting and proliferating subpopulations grow exponentially with rate $\lambda$. If $\lambda<0$, the proliferating subpopulation becomes extinct and leaves behind a pool of persisting cells.
}

\newpage
\begin{center}
	\Large\textbf{Supplemental figures}
\end{center}

\begin{figure*}[h!]
	\includegraphics[width=17.5cm]{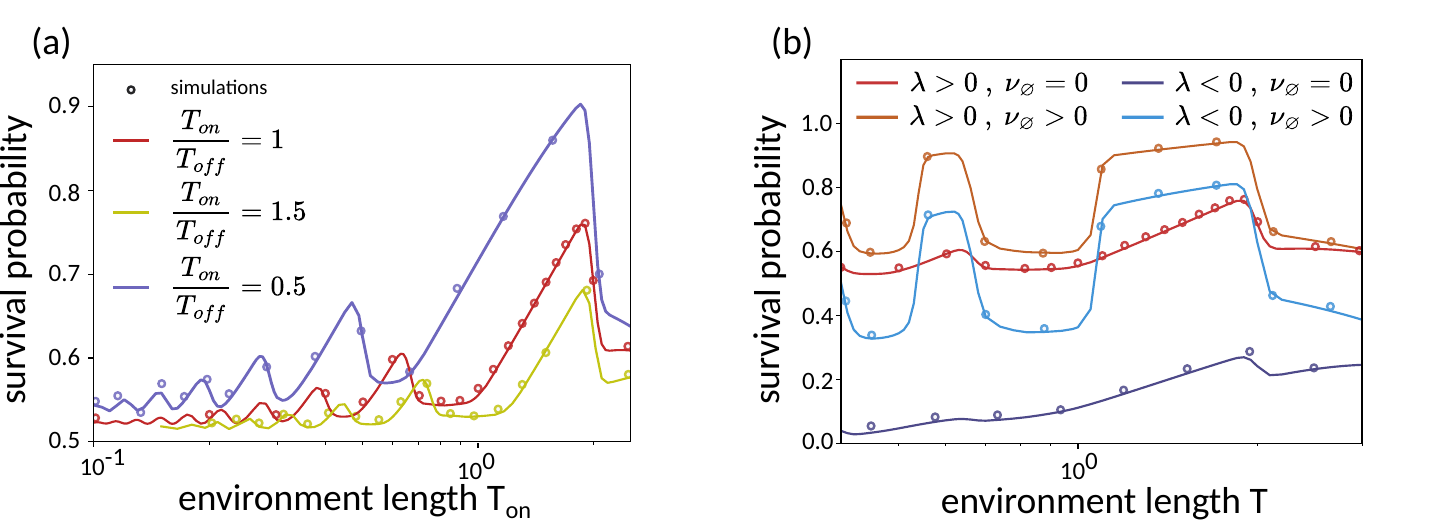}
	\caption{
		%add textbf title for each suppl. figs
		\textbf{Dependence of noise-induced resonances on environment shape, growth regimes and persistence probability.}
		(a) Survival resonances as a function of the on-environment length ($T_\text{on}$) for pulsed ($\frac{T_\text{on}}{T_\text{off}}=0.5$, purple line), even ($\frac{T_\text{on}}{T_\text{off}}=1$, red line), and anti-pulsed ($\frac{T_\text{on}}{T_\text{off}}=1.5$, yellow line) modulations of the on-environment. The survival probability is peaked around $T_\text{on}=\mu_d-nP$ with period $P=T_\text{on}+T_\text{off}$, in agreement with Eq.~\eqref{eq_noise-induced_resonance}. (b) Noise-induced resonances exist both in sub-($\lambda<0$ during the on-environment, blue line)  and supercritical ($\lambda>0$ during the on-environment, red line) growth regimes. The amplification induced by persistence exists both in sub-($\lambda<0$ during the on-environment, light blue line)  and supercritical ($\lambda>0$ during the on-environment, orange line) growth regimes. Division and death times are $\Gamma$ distributed with parameters $\md=2$, $\cvd=10^{-3}$, $\mg=1.8$, $\cvg=1$, $w_\gamma=w_d=1$ in (a); $\md=2$, $\cvd=10^{-3}$, $\mg=1.8$, $\cvg=1$ (red line $w_\gamma=w_d=1$, orange line $w_\gamma=w_d=0.99$) and $\mg=2.94$ (blue line $w_\gamma=w_d=1$, light blue line $w_\gamma=w_d=0.99$) in (b).}
\label{sfig:robustness}
\end{figure*}

\begin{figure*}[h!]
\includegraphics[width=8.5cm]{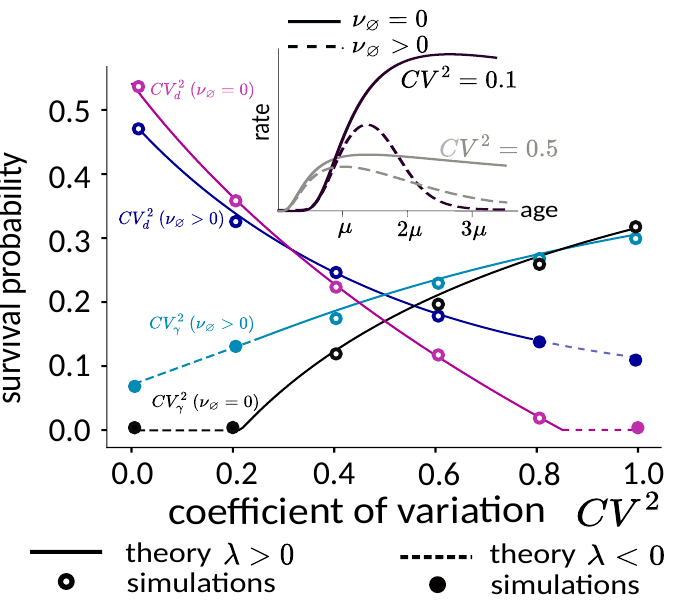}
\caption{\textbf{Survival probability in a constant environment with log-normal distributed division and death times.} Survival probability shows transitions from extinction to fractional killing as a function of the coefficient of variations in division or death times, $\cvg$ and $\cvd$. The transitions from decaying ($\lambda<0$, dashed) to growing dynamics ($\lambda>0$, solid lines) are indicated through points. Log-normal-distributed division and death times with $\md=1$, $\mg=0.9$, $\cvd=\cvg=0.5$ an $w_{\g}=w_d=0.8$ (corresponding to $\nu_\emptyset=0.04$) are assumed.}
\label{sfig:lognormal}
\end{figure*}

\end{document}